\documentclass[11pt,twoside]{article}

\usepackage{asp2006}
\usepackage{lscape}
\usepackage{graphicx}

\def\eq#1{\begin{equation} #1 \end{equation}}
\def\E#1{\hbox{$10^{#1}$}}
\def\about {\hbox{$\sim$}}
\def\x     {\hbox{$\times$}}
\def\mic   {\hbox{$\mu$m}}
\def\deg   {\hbox{$^\circ$}}
\def\tV    {\hbox{$\tau_{\rm V}$}}

\def\No    {\hbox{${\cal N}_0$}}

\def\Rd     {\hbox{$R_{\rm d}$}}
\def\Ro     {\hbox{$R_{\rm o}$}}
\def\Rc     {\hbox{$R_{\rm c}$}}
\def\Rx     {\hbox{$R_{\rm x}$}}
\def\erg   {\hbox{erg\,s$^{-1}$}}

\def\MBH   {\hbox{$M_{\bullet\,7}$}}
\def\rpc   {\hbox{$r_{\rm pc}$}}
\def\Mo     {\hbox{$M_{\odot}$}}

\def\Nc     {\hbox{$N_{\rm C}$}}

\def\NH     {\hbox{$N_{\rm H}$}}

\def\cs    {\hbox{cm$^{-2}$}}
\def\Macc  {\hbox{$\dot M_{\rm acc}$}}

\makeatletter
\def\@jourvol{***}
\def\cpr@year{Xi'an, China, 16--21 October, 2006}
\def\vol@title{The Central Engine of Active Galactic Nuclei}
\def\vol@author{eds. L. C. Ho and J.-M. Wang}
\makeatother

\begin{document}

\title{Unification Issues and the AGN Torus}

\author{Moshe Elitzur}
\affil{Physics \& Astronomy Dept. \\
       University of Kentucky     \\
       Lexington, KY 40506-0055   \\
       USA}

\begin{abstract}
Observations give strong support for the unification scheme of active galactic
nuclei. Clumpiness of the toroidal obscuration is crucial for explaining the IR
observations and has significance consequences for AGN classification: type 1
and type 2 viewing is an angle-dependent probability, not an absolute property.
The broad line region (BLR) and the dusty torus are, respectively, the inner
and outer segments, across the dust sublimation radius, of a continuous cloud
distribution. Continuum X-ray obscuration comes mostly from the inner, BLR
clouds. All clouds are embedded in a disk wind, whose mass outflow rate is
diminishing as the accretion rate, i.e., AGN luminosity, is decreasing. The
torus disappears when $L \la$ \E{42} \erg, the BLR at some lower, yet to be
determined luminosities.

\end{abstract}

\section{Introduction}

The basic premise of the unification scheme is that every AGN is intrinsically
the same object: an accreting supermassive black hole. This central engine is
surrounded by a dusty toroidal structure so that the observed diversity simply
reflects different viewing angles of an axisymmetric geometry. Since the torus
provides anisotropic obscuration of the center, sources viewed face-on are
recognized as ``type 1'', those observed edge-on are ``type 2''.

A scientific theory must make falsifiable predictions, and AGN unification does
meet this criterion. Unification implies that for every class of type 1 objects
there is a corresponding type 2 class, therefore the theory predicts that type
2 QSO must exist. After many years of searching, QSO2 have been discovered,
thanks to the {\em Sloan Digital Sky Survey}. Furthermore, spectro\-polarimetry
of type~2 quasars even reveals the hidden type 1 nuclei at $z$ as large as 0.6
(Zakamska et al 2005). This is a spectacular success of the unification
approach. There are not that many cases in astronomy --- in fact, in all of
science --- where a new type of object has been predicted to exist and then
actually discovered.

In light of this success, it would be hard to question the basic validity of
the unification approach. There is no reason, though, why the obscuring torus
should be the same in every AGN; it is unrealistic to expect AGN's to differ
only in their overall luminosity but be identical in all other aspects. Here I
summarize the properties of the obscuring torus and try to speculate on how it
might evolve with the AGN luminosity, i.e., its accretion rate.

\section{IR Emission}

Obscuring dust must re-radiate the absorbed radiation at longer wavelengths.
Indeed, type 1 AGN display at short wavelengths (X-rays through optical) the
power law spectrum characteristic of accretion disks and at $\lambda \ga$
1\mic\ an IR bump, which can be attributed to reprocessing by the torus dust
(Barvainins 1987); a striking example is provided by the recent observations of
Mrk 1239 (Rodr{\'{\i}}guez-Ardila \& Mazzalay \ 2006). In contrast, type 2
sources display only the IR emission, as expected. The SED's do conform to the
unification prediction: type 1 = type 2 + AGN.

IR observations also bring puzzles. The 10 \mic\ torus emission in NGC1068 was
recently resolved in VLTI interferometry by Jaffe et al (2004). They analyzed
their observations with a model that contains a compact ($r \la$ 0.5 pc) hot
($>$ 800 K) core and cooler (320 K) dust extending to $r \simeq$ 1.7 pc.
Poncelet et al (2006) reanalyzed the same data with slightly different
assumptions and reached similar conclusions --- the coolest component in their
model has an average temperature of 226 K and extends to $r$ = 2.7 pc. As noted
by the latter authors, the presence of dust temperatures of only 200--300 K so
close to the AGN is a most puzzling, fundamental problem. The bolometric
luminosity of NGC1068 is \about\ 2\x\E{45} erg s$^{-1}$ (Mason et al 2006),
therefore the dust temperature expected at $r$ = 2 pc is 960 K, much higher
than those found by the observations; $T$ = 320 K is expected only at $r$ = 26
pc, not the 1.7 pc that Jaffe et al find, and $T$ = 226 K should be still
further out at $r$ = 57 pc, not the 2.7 pc that Poncelet et al find.

%%%%%%%%%%%%%%% Tm-r %%%%%%%%%%%%%%%%%%%%%%%%%%
\begin{figure}[ht]
 \centering\leavevmode\includegraphics[width=0.8\hsize,clip]{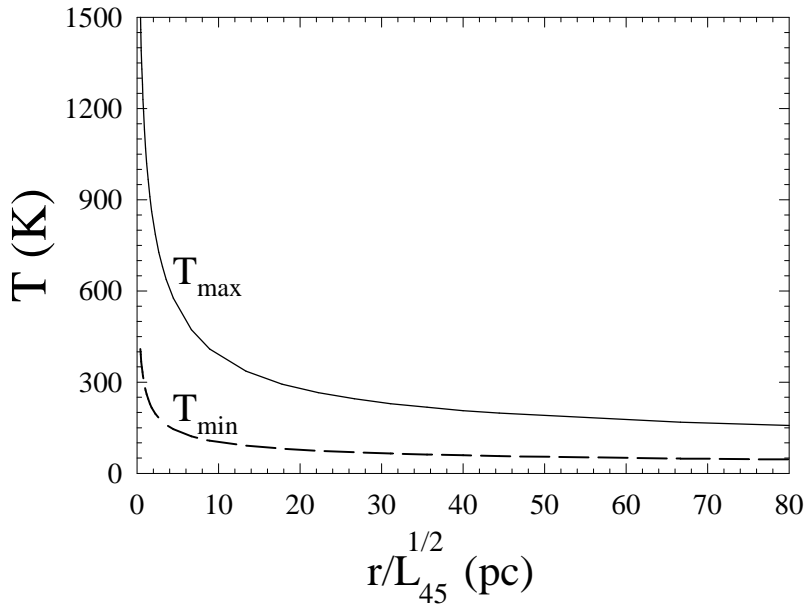}
\caption{The highest ($T_{\rm max}$) and lowest ($T_{\rm min}$) dust
temperatures on the surface of an optically thick cloud at distance $r$ from an
AGN with luminosity $L_{45} = L$/(\E{45} erg\,s$^{-1}$). The highest
temperature occurs on the illuminated face, the lowest on the dark side
(Nenkova et al 2006).} \label{Fig:Tm-r}
\end{figure}
%%%%%%%%%%%%%%%%%%%%%%%%%%%%%%%%%%%%%%%%%%%%%%%%%%%%%%%

These discrepancies are resolved when we recall that the torus is comprised of
dusty clouds, which are individually optically thick (Krolik \& Begelman 1988).
The temperature of an optically thick dusty cloud is much higher on the side
illuminated by the AGN than on the opposite, dark side (Nenkova et al 2002).
Figure \ref{Fig:Tm-r} displays the variation of the surface temperatures on the
bright side ($T_{\rm max}$) and dark side ($T_{\rm min}$) of an optically thick
cloud with distance from the AGN. While the dust temperature on the bright
sides of clouds at \about\ 2 pc from the center of NGC1068 is 950 K, the
temperature on their dark sides is only 250 K, declining to 210 K at 3 pc.
Indeed, the temperatures deduced in the model synthesis of the VLTI data fall
inside the range covered by the cloud surface temperatures at the derived
distances.

Another puzzle is the apparent similarity between the IR emission from type 1
and type 2 sources: the torus obscuration is highly anisotropic (which is the
essence of unification), yet its emission seems to be nearly isotropic. This
was first noted by Lutz et al (2004). They compared the 6 \mic\ emission of
Seyfert 1 and 2 galaxies normalized to the intrinsic hard X-ray emission, and
concluded that the distributions of the two populations are essentially
identical within the observational errors. Horst et al (2006) used the same
approach for the 12 \mic\ emission and reached similar conclusions. Buchanan et
al (2006) conducted Spitzer observations of 87 Seyfert galaxies in the
$\lambda$ = 5--35 \mic\ range and normalized the IR fluxes with the optically
thin radio emission, which is free from the large corrections required to
determine the intrinsic X-ray flux densities from the observed ones. Although
at 6 \mic\ they find a larger variation than Lutz et al, they find that the
emission from Seyfert 1 and 2 galaxies are within factor 2 of each other for
all $\lambda \ga$ 15 \mic. The authors of all these studies note the problems
the observations pose to smooth-density torus models, which predict a highly
anisotropic torus emission.

\begin{figure}
 \centering\leavevmode\includegraphics[width=0.7\hsize,clip]{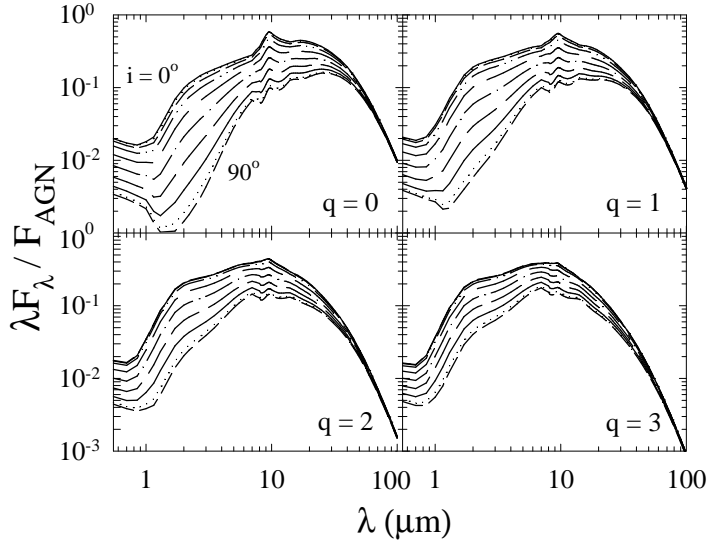}
\caption{Model SED's of clumpy dusty tori. There are \No\ = 5 clouds, on
average, along radial equatorial rays, each with optical depth \tV\ = 60 at
visual. The clouds angular distribution is Gaussian with $\sigma$ = 45\deg, the
radial distribution is a power law (see eq.\ \ref{eq:Nc}). Each panel shows a
different power index $q$, as marked. The different curves in each panel
correspond to different viewing angles $i$, varying from 0\deg\ to 90\deg\ in
10\deg\ steps (Nenkova et al 2006).} \label{Fig:SED-qdep}
\end{figure}

This problem, too, is solved by the torus clumpiness. Nenkova et al (2006)
performed detailed radiative transfer calculations for the emission from a
clumpy torus in which each cloud is characterized by \tV, its optical depth at
visual. The cloud distribution starts at the dust sublimation radius, and the
number of clouds encountered on average along radial equatorial rays is \No.
The cloud angular distribution is parametrized as Gaussian with width parameter
$\sigma$, the radial distribution as an inverse power law with index $q$; that
is, the number of clouds per unit length as a function of radial distance $r$
and angle $\beta$ from the equatorial plane is\footnote{The proportionality
constant is determined from $\No = \int\Nc(r,\beta = 0) dr$}
\eq{\label{eq:Nc}
    \Nc(r,\beta) \propto \No e^{(-\beta/\sigma)^2}r^{-q}
}
Figure \ref{Fig:SED-qdep} shows sample SED's for models in which all the
parameters are the same except for $q$, the index of the power-law radial
distribution. At any viewing angle, the AGN obscuration is identical for all
these models since obscuration depends only on the total number of clouds along
the radial ray, which is the same in all cases. Because of the Gaussian angular
distribution, the obscuration is highly anisotropic, decreasing by two orders
of magnitudes between polar and equatorial viewing. However, the variation of
SED with viewing angle is quite moderate and the emission becomes progressively
more isotropic as $q$ increases (the radial distribution gets steeper). A
clumpy torus can produce extremely anisotropic obscuration of the AGN together
with nearly isotropic IR emission, as observed. In particular, the mild
anisotropy of the $q$ = 2 models is compatible with that found by Buchanan et
al.

%%%%%%%%%%%%%%%%%%%%%%%%%%%%%%%%%%%%%%%%%%%%%%%%%%%%%%
%%%%%%%%%%%%%%%%%%%%%%%%%%%%%%%%%%%%%%%%%%%%%%%%%%%%%%%

All in all, clumpy torus models (Nenkova et al 2002, 2006) seem to produce
SED's that are in reasonable overall agreement with observations for the
following range of parameters:
\begin{itemize}
\item
Number of clouds, on average, along radial equatorial rays \No\ = 5--10

\item
Angular width $\sigma$ = 30\deg--60\deg

\item
Index of power law radial distribution $q$ = 1--2

\item
Optical depth, at visual, of each cloud \tV\ = 40--120

\item
Torus extends from dust sublimation at \Rd\ = 0.4$L_{45}^{1/2}$ pc to an outer
radius \Ro\ $\ge$ 5\Rd

\end{itemize}
There is increasing evidence that the torus is quite compact (see Elitzur 2006
and references therein). All observations are consistent with \Ro/\Rd\ being no
larger than \about\ 20--30, and perhaps even only \about\ 5--10. It is
especially significant that, thanks to the low temperatures on the dark sides
of nearby clouds, models of clumpy tori can produce sufficient IR emission with
dimensions that are rather compact.

While the above listed parameters lead to SED's compatible with observations,
some additional considerations can further restrict the acceptable range. For
example, the near-isotropy of IR emission indicates that $q$ = 2 may provide a
more appropriate description of the radial distribution than $q$ = 1. Finally,
all model calculations were performed with standard Galactic ISM dust, which
seems to provide satisfactory results. Current data do not provide any
compelling reason for drastic changes in the dust composition.

\section{Clumpy Unification}

\begin{figure}
 \centering\leavevmode
 \includegraphics[width=0.45\hsize,clip]{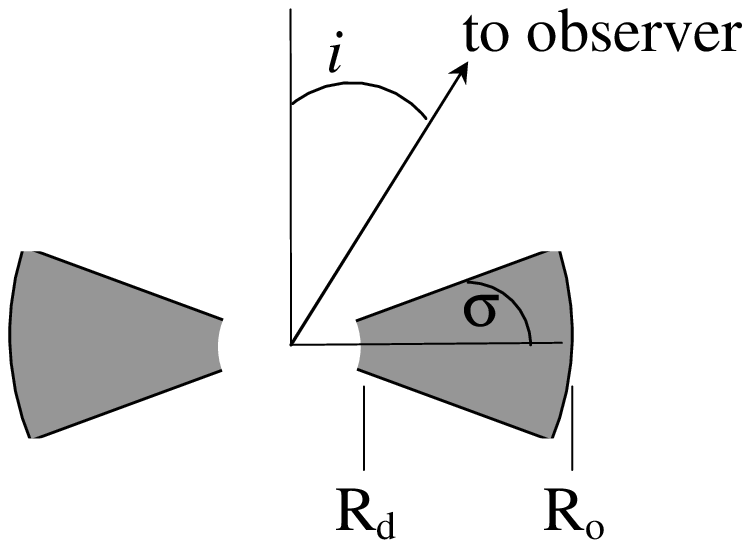} \hfill
 \includegraphics[width=0.45\hsize,clip]{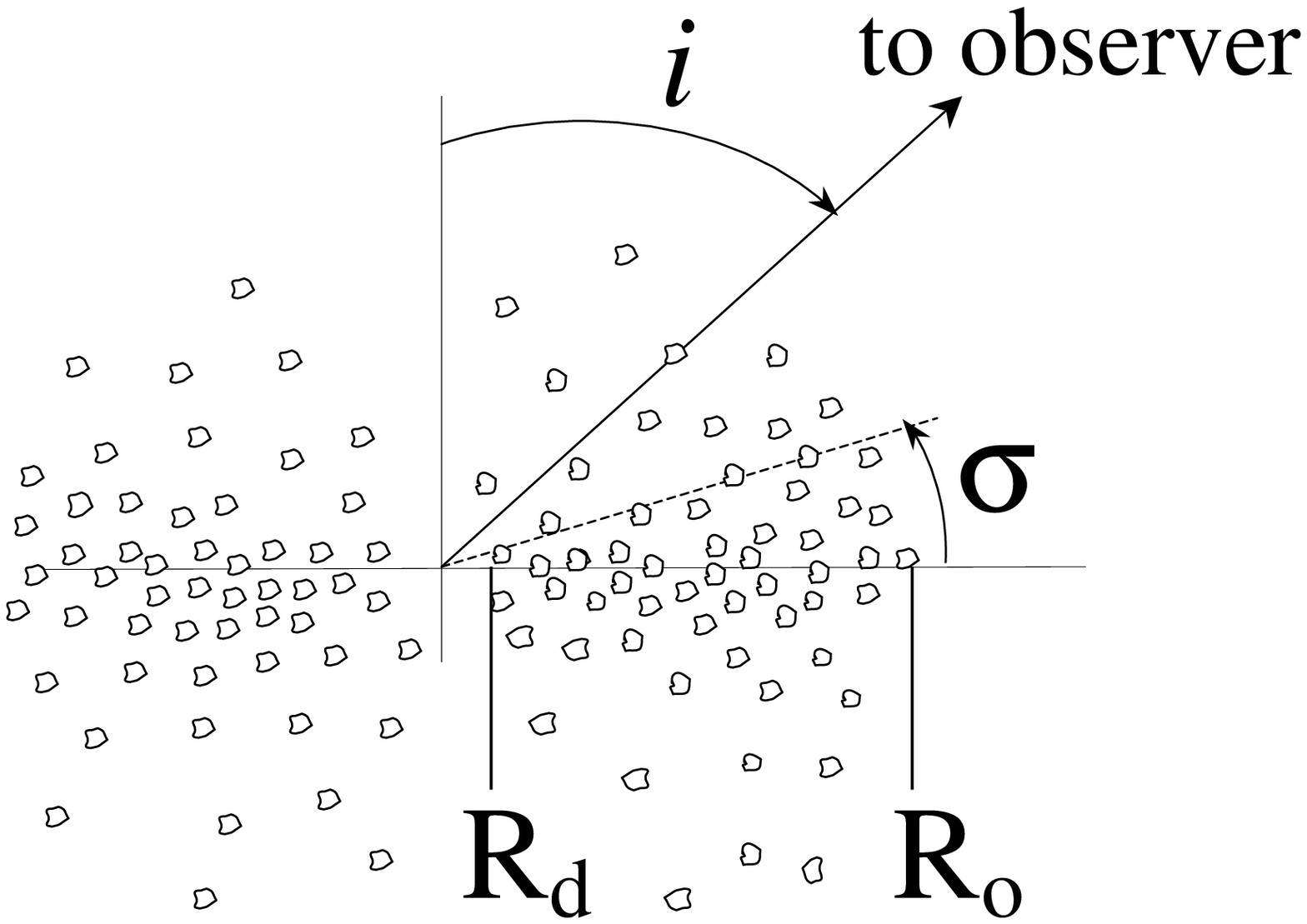}
\caption{AGN classification according to unified schemes. {\em Left}: In a
smooth-density torus, the viewing angle $i = \frac12\pi - \sigma$ separates
between type 1 and type 2 viewing. {\em Right:} In a clumpy, soft-edge torus,
the probability for direct viewing of the AGN decreases away from the axis, but
is always finite.} \label{Fig:Smooth_Clumpy}
\end{figure}

The classification of AGN into types 1 and 2 is based on the extent to which
the nuclear region is visible. In its standard formulation, the unification
approach posits the viewing angle as the sole factor in determining the AGN
type. This is indeed the case for a smooth-density torus that is optically
thick within the angular width $\sigma$ (figure \ref{Fig:Smooth_Clumpy}, left
sketch). All AGN viewed at $0 \le i < \sigma$ then appear as type 1 sources,
while viewing at $\sigma \le i \le \frac12\pi$ gives type 2 appearance. If
$f_2$ denotes the fraction of type 2 sources in the total population, then $f_2
= \sin\sigma$. From statistics of Seyfert galaxies Schmitt et al (2001) find
that $f_2 \simeq$ 70\%, hence their estimate $\sigma \simeq$ 45\deg. The issue
is currently unsettled because Hao at al (2005) have found recently that $f_2$
is only about 50\%, or $\sigma \simeq$ 30\deg.

%%%%%%%%%%%%%%%%%%%%%%%%%%%%%%%%%%%%%%%%%%%%%%%%%%%%%%
%%%%%%%%%%%%%%%%%%%%%%%%%%%%%%%%%%%%%%%%%%%%%%%%%%%%%%%

Because of clumpiness, the difference between types 1 and 2 is not truly an
issue of orientation but of probability for direct view of the AGN (figure
\ref{Fig:Smooth_Clumpy}, right sketch); {\em AGN type is a viewing-dependent
probability.} Since that probability is always finite, type 1 sources can be
detected from what are typically considered type 2 orientations, even through
the torus equatorial plane: if \No\ = 5, for example, the probability for that
is $e^{-5} = 1/148$ on average. This might offer an explanation for the few
Seyfert galaxies reported by Alonso-Herrero et al (2003) to show type 1 optical
line spectra together with 0.4--16 \mic\ SED that resemble type 2. Conversely,
if a cloud happens to obscure the AGN from an observer, that object would be
classified as type 2 irrespective of the viewing angle. In cases of such single
cloud obscuration, on occasion the cloud may move out of the line-of-sight,
creating a clear path to the nucleus and a transition to type 1 spectrum. Such
transitions between type 1 and type 2 line spectra have been observed in a few
sources (see Aretxaga et al 1999, and references therein).  It is worth while
to conduct monitoring observations in an attempt to detect additional such
transitions. The most promising candidates would be obscured systems with
relatively small X-ray obscuring columns, small torus sizes and large
black-hole masses (Nenkova et al 2006).

%%%%%%%%%%%%%%%%%%%%%%%%%%%%%%%%%%%%%%%%%%%%%%%%%%%%%%%
\begin{figure}
% \Figure{type2fraction}{\figsize}
 \centering\leavevmode\includegraphics[width=0.7\hsize,clip]{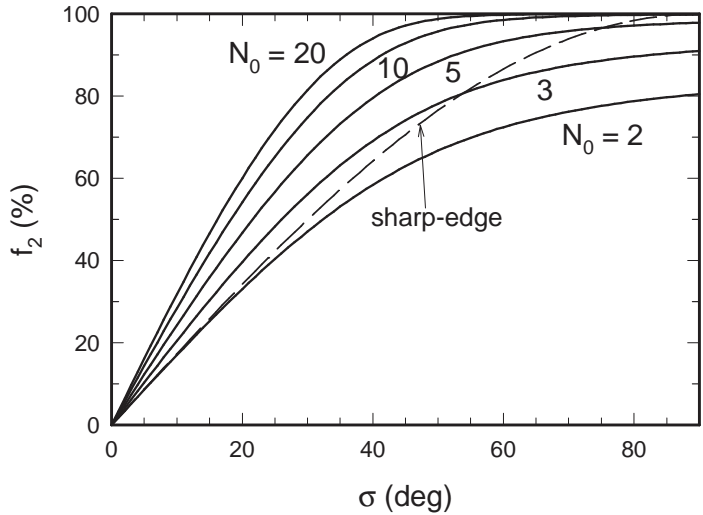}
\caption{AGN statistics: The fraction $f_2$ of obscured sources as a function
of the torus width parameter $\sigma$. In a uniform density sharp-edge torus
(see fig.\ \ref{Fig:Smooth_Clumpy}), this fraction is determined uniquely by
$\sigma$, and is shown with the dashed line. In contrast, in a clumpy torus
with Gaussian angular distribution (eq.\ \ref{eq:Nc}), $f_2$ depends on both
$\sigma$ and the cloud number \No, which is marked on the various solid lines
(Nenkova et al 2006).
} \label{Fig:f2}
\end{figure}
%%%%%%%%%%%%%%%%%%%%%%%%%%%%%%%%%%%%%%%%%%%%%%%%%%%%%%%

A sharp-edge clumpy torus has $f_2 = (1 - e^{-{\cal N}_0})\sin\sigma$, and is
practically indistinguishable from its smooth-density counterpart when $\No$
exceeds \about\ 3--4. However, the situation changes fundamentally for
soft-edge distributions because at every viewing angle, the probability of
obscuration increases with the number of clouds. As is evident from figure
\ref{Fig:f2}, the Gaussian distribution produces a strong dependence on \No\
and significant differences from the sharp-edge case. Since the sharp-edge
angular distribution is ruled out by observations (Nenkova et al 2006), {\em
the fraction of obscured sources depends not only on the torus angular width
but also on the average number of clouds along radial rays}. While the fraction
$f_2 = 70\%$ requires $\sigma = 45\deg$ in the smooth-density case, it implies
$\sigma = 33\deg$ in a Gaussian clumpy torus with \No\ = 5 clouds.

There are indications that the fraction $f_2$ of obscured sources decreases
with luminosity (Hao et al 2005; Simpson 2005), although counter claims exist
too (see talk by Junxian Wang, these proceedings). The possible decrease of
$f_2$ with $L$ has been interpreted as support for the ``receding torus''
model, in which $\sigma$ decreases with $L$ (Simpson 2005 and references
therein). However, all the quantitative analyses performed thus far for the
$L$-dependence of $f_2$ were based on sharp-edge angular obscuration. Removing
this assumption affects profoundly the foundation of the receding torus model
because the dependence on the number of clouds necessitates analysis with two
free parameters, therefore $\sigma$ cannot be determined without \No. A
decrease of \No\ with $L$ at constant $\sigma$ will also produce a decrease in
$f_2$, the same effect as a decrease of $\sigma$ (figure \ref{Fig:f2}). An
observed trend of $f_2$ with $L$ may arise from a dependence on either $\sigma$
or \No\ or both. There is no obvious a-priori means for deciding between the
various possibilities.

\section{X-rays and Unification}

X-ray observations give overwhelming evidence for the orientation-dependent
absorption and reprocessing expected from AGN unification (see talk by R.\
Maiolino, these proceedings). But in spite of the close correspondence between
the two, the ``X-ray torus'' probably does not coincide with the ``dusty
torus''. The decisive evidence comes from the short time scales for transit of
X-ray absorbing clouds across the line of sight, which establish the clouds
location inside the dust sublimation radius. A recent extreme case is the
two-day flips between Compton thick (\NH\ $>$ \E{24} \cs) and thin X-ray
absorption in NGC 1365 (G.\ Risaliti, these proceedings). These observations
show that the torus extends inward of the dust sublimation point to some inner
radius \Rx\ $<$ \Rd. Clouds at $\Rx \le r \le \Rd$ partake in X-ray absorption
but do not contribute appreciably to optical obscuration or IR emission because
they are dust-free. {\em The bulk of the X-ray absorption likely comes in most
cases from the clouds in the dust-free inner portion of the torus}. The columns
for X-ray absorption exceed those for UV absorption (Maiolino et al 2001),
showing that the inner radial segments $\Rx \le r \le \Rd$ contain more clouds
than the outer segments at $r > \Rd$. This is the expected behavior in the
steep $1/r^2$ distribution, where most of the clouds are concentrated toward
the center. All observations are consistent with \Rx\ \about\ 0.1\Rd\ and
roughly 10 dust-free X-ray absorbing clouds for every dusty cloud, therefore
the fraction of absorbed sources should be higher in X-rays than in UV/optical.
This explains the discovery of type 1 QSO that show Compton thick X-ray
absorption but no UV absorption (Gallagher et al 2006; see also talk by S.\
Gallagher, these proceedings).

\section{What is the Torus?}

In the ubiquitous sketch by Urry \& Padovani (1995), the AGN central region,
which is comprised of the black hole, its accretion disk and the broad-line
emitting clouds, is surrounded by a large doughnut-like structure --- the
torus. This hydrostatic object is a separate entity, presumably populated by
molecular clouds accreted from the galaxy. Gravity controls the orbital motions
of the clouds, but the origin of vertical motions capable of sustaining the
``doughnut" as a steady-state, hydrostatic structure whose hight is comparable
to its radius (for $\sigma$ \about\ 45\deg) was recognized as a problem since
the first theoretical study by Krolik \& Begelman (1988). This problem has
eluded solution to this date.

Two different types of observations now show that the torus is in fact a smooth
continuation of the broad lines region, not a separate entity. The first
evidence comes from the IR reverberation results of Suganuma et al (2006). They
show that the dust sublimation radius scales with luminosity as $L^{1/2}$, as
expected. More importantly, the sizes of broad line emission regions scale
similarly as $L^{1/2}$ and in each source for which both data exist, the IR
time lag is the upper bound on all time lags measured in the broad lines; that
is, the BLR extends all the way to the inner boundary of the dusty torus, a
relation verified over a range of \E6\ in luminosity. This finding validates
the Netzer \& Laor (1993) proposal that the BLR size is bounded by dust
sublimation. The second evidence is the finding by Risaliti et al (2002) that
the X-ray absorbing columns in Seyfert 2 display time variations caused by
cloud transit across the line of sight. Most variations come from clouds that
are dust free because of their proximity ($<$ 0.1 pc) to the AGN, but some
involve dusty clouds at a few pc. Other than the different time scales for
variability, there is no discernible difference between the dust-free and the
dusty X-ray absorbing clouds.

These observations show that the X-ray absorption, broad line emission and dust
obscuration and reprocessing are produced by a single, continuous distribution
of clouds. The different radiative signatures merely reflect the change in
cloud composition across the dust sublimation radius \Rd. The inner clouds are
dust free. Their gas is directly exposed to the AGN ionizing continuum,
therefore it is atomic and ionized, producing the broad emission lines. The
outer clouds are dusty, therefore their gas is shielded from the ionizing
radiation and the atomic line emission is quenched. Instead, the material in
these clouds is molecular and dusty, obscuring the optical/UV emission from the
inner regions and emitting IR. Thus the BLR occupies $r \le \Rd$ while the
torus is simply the $r > \Rd$ region. Both regions produce X-ray absorption,
but because each radial ray contains most of its clouds in its BLR segment,
that is where the bulk of the X-ray obscuration is produced. Since the X-ray
obscuration region coincides mostly the with BLR, it seems appropriate to name
this region instead BLR/XOR. By the same token, since the unification torus is
just the outer portion of the cloud distribution and not an independent
structure, it is appropriate to rename it the TOR for Toroidal Obscuration
Region.

The merger of the ionized and dusty clouds into a single population offers a
solution to the torus vertical structure problem. Mounting evidence for cloud
outflow (see, e.g., Elvis 2004) indicates that instead of a hydrostatic
``doughnut'', the TOR is a region in a clumpy wind coming off the accretion
disk (see Elitzur \& Shlosman 2006 and references therein). The accretion disk
appears to be fed by a midplane influx of cold, clumpy material from the main
body of the galaxy. The outer regions of the disk are dusty and molecular, as
observed in water masers in some edge-on cases. At smaller radii the disk
composition switches to atomic and ionized, producing a double-peak signature
in emission line profiles. Approaching the center, conditions for developing
hydromagnetically- or radiatively-driven winds above this equatorial inflow
become more favorable. The disk-wind geometry provides a natural channel for
angular momentum outflow from the disk and is found on many spatial scales,
from protostars to AGN. In both the inner (atomic and ionized) and outer (dusty
and molecular) regions, as the clouds rise and move away from the disk they
expand and lose their column density. Therefore the X-ray absorption, broad
line emission and dust obscuration and emission all have a limited vertical
scope, resulting in a toroidal geometry for both the BLR/XOR and the TOR.

%%%%%%%%%%%%%%% AGN Scheme %%%%%%%%%%%%%%%%%%%%%%%%%%
\begin{figure}[ht]
 \centering\leavevmode\includegraphics[width=0.6\hsize,clip]{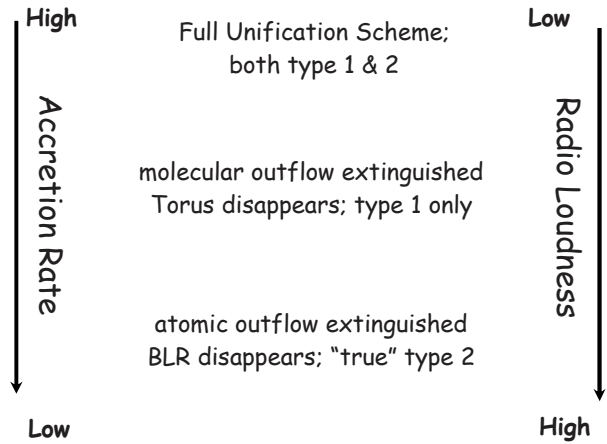}
\caption{Conjectured scheme for AGN evolution with decreasing accretion rate.}
\label{Fig:AGN_scheme}
\end{figure}
%%%%%%%%%%%%%%%%%%%%%%%%%%%%%%%%%%%%%%%%%%%%%%%%%%%%%%%

The properties of the dusty TOR clouds are determined from their IR emission
and the requirement that they withstand tidal shearing. The cloud column
density is \NH\ \about\ \E{22} -- \E{23} cm$^{-2}$. At a distance \rpc\ (in pc)
from a black hole with mass \E7\MBH\ (in \Mo), the cloud density is $n >$ \E7\
\MBH/$\rpc^3$  cm$^{-3}$, its size is \Rc\ $<$ \E{16}  \NH$_{23} \rpc^3$/\MBH\
cm and its mass $M_c < 7\cdot\E{-3}$ \NH$_{23}\Rc_{16}^2$ \Mo. These parameters
are similar to what is found at molecular cores of ISM clouds. Indeed,
molecular clouds with such properties and uplifted from the disk appear to have
been detected in water maser observations of NGC 3079 (Kondratko et al.\ 2005).

A key prediction of the wind scenario is that the TOR disappears at low
bolometric luminosities ($\la$ \E{42} \erg) because mass accretion can no
longer sustain the required cloud outflow rate. This prediction seems to be
corroborated in observations of both FR I radio galaxies (Chiaberge et al.\
1999) and LINERs (Maoz et al 2005). In particular, the histogram of UV colors
shows an overlap between the two populations of type 1 and type 2 LINERs with
$L \la$ \E{42} \erg. The difference between the peaks corresponds to dust
obscuration in the type 2 LINERs of only \about\ 1 magnitude in the R band,
minute in comparison with higher luminosity AGN. If the TOR was the only
component removed from the system, all low luminosity AGN would become type 1
sources. In fact, among the low-obscuration LINERs Maoz et al (2005) find
sources both with broad H$\alpha$ wings (type 1) and without (type 2).
Therefore the broad line component is truly missing in the type 2 sources in
this sample. Similarly, Laor (2003) presents arguments that some ``true'' type
2 sources, i.e., having no obscured BLR, do exist among AGN with $L \la
\E{42}\,\erg$. Both findings have a simple explanation if when $L$ decreases
still further, the suppression of mass outflow spreads radially inward from the
disk's dusty, molecular region into its atomic, ionized zone. Then the TOR
disappearance would be followed by a diminished outflow from the inner ionized
zone and disappearance of the BLR/XOR at lower, still to be determined
luminosities. This evolutionary scheme is sketched in figure
\ref{Fig:AGN_scheme}.

Ho (2002) finds that the radio loudness of AGN is {\em inversely} correlated
with the mass accretion rate. That is, when \Macc\ is decreasing, the cloud
outflow rate is decreasing too while the radio loudness is increasing. It seems
that the AGN switches its main dynamic channel for release of excess accreted
mass from torus outflow at higher luminosities to radio jets at lower ones.
X-ray binaries display a similar behavior, switching between radio quiet states
of high X-ray emission and radio loud states with low X-ray emission.
Comparative studies of AGN and X-ray binaries seem to be a most useful avenue
to pursue.

\acknowledgements Partial support by NASA and NSF is gratefully acknowledged.

\end{document}